\documentclass[12pt, draftclsnofoot, journal, letterpaper, onecolumn]{IEEEtran}
\usepackage[cmex10]{amsmath}
\usepackage{algorithmic}
\usepackage{array}
\usepackage[caption=false,font=footnotesize]{subfig}
\usepackage{fixltx2e}
\usepackage{stfloats}
\usepackage{mathrsfs}
\usepackage{amsmath}
\usepackage{amssymb}
\usepackage{bbm}
\usepackage{url}
\usepackage{xcolor}
\usepackage{cleveref}
\usepackage{graphicx}
\usepackage{epstopdf}
\usepackage{cite}
\newtheorem{lemma}{Lemma}[section]
\newtheorem{corollary}{Corollary}[section]
\newtheorem{theorem}{Theorem}[section]

\begin{document}

\title{Power Efficiency for Device-to-Device Communications}

\author{\IEEEauthorblockN{Yanbo~Ma\IEEEauthorrefmark{1}, %~\IEEEmembership{Member,~IEEE,}
        Yuan~Liu\IEEEauthorrefmark{1}, %~\IEEEmembership{Member,~IEEE,}
        and~Meixia~Tao\IEEEauthorrefmark{1}}\\%,~\IEEEmembership{Senior~Member,~IEEE}
\IEEEauthorblockA{\IEEEauthorrefmark{1}Department of Electronic Engineering, Shanghai Jiao Tong University, Shanghai, P. R. China}\\
Email: myb800@sjtu.edu.cn, \{eeyuanliu, mxtao\}@ieee.org
}

\maketitle

\begin{abstract}
The concept of device-to-Device (D2D) communication as an underlay coexistence with cellular networks gains many advantages of improving system performance.
In this paper, we model such a two-layer heterogenous network based on stochastic geometry approach. We aim at minimizing the expected power consumption of the D2D layer while satisfying the outage performance of both D2D layer and cellular layer. We consider two kinds of power control schemes. The first one is referred as to \emph{independent power control} where the transmit powers are statistically independent of the networks and all channel conditions. The second is named as \emph{dependent power control} where the transmit power of each user is dependent on its own channel condition. A closed-form expression of optimal independent power control is derived, and we point out that the optimal power control for this case is fixed and not relevant to the randomness of the network. For the dependent power control case, we propose an efficient way to find the close-to-optimal solution for the power-efficiency optimization problem.
Numerical results show that dependent power control scheme saves about half of power that the independent power control scheme demands.
\end{abstract}


\IEEEpeerreviewmaketitle

\section{Introduction}

The exponentially increasing data traffic and requirements of user experiences call for dramatic expansion of energy consumption. To meet the high demands of resource saving and environment protection for wireless networks today, it is quite necessary and urgent to save energy consumption from network nodes, such as base stations, access points, and mobile devices.

Device-to-Device (D2D) communication is a promising technology and allowed as an underlay coexistence with cellular networks. It enables a pair of devices in proximity of each other to establish a direct local link and is not through a base station or access point. Such a heterogenous network infrastructure has attracted much attention due to its potential of improving system performance, such as throughput enhancing, coverage extension, and data offloading \cite{D2DLTEA,DAD2D,PCD2D}.

Stochastic geometry \cite{baccelli2009stochastic,baccelli2009stochastic2,interference} is a powerful tool that provides tractable analysis for large wireless networks \cite{overlaid,SGRG,RPC2,FPC,OP}. For instance, \cite{overlaid} investigates feasibility region about node density. The work in \cite{SGRG} summarizes fundamental limits of wireless cellular network based on stochastic geometry, like connectivity, capacity, outage, etc.
The authors in \cite{RPC2} consider power control in random networks with Poisson distributed nodes using a game theoretic approach. Moreover, \cite{RPC2} adopts ALOHA-type random on-off power control policies to maximize expected local performance of each link. The authors in \cite{OP} present the channel inversion based power control for an ad hoc network. The study of \cite{FPC} propose the fractional power control for a single homogeneous network, where the power control is the fractional exponent of the channel.

In this paper, we consider a large scale network where D2D communication reuse the uplink spectrum of the cellular communication. Using stochastic geometry, we model the network nodes as Poisson Point Processes (PPP). Our goal is to study the feasibility of such a two-layer network with power optimization first. Based on the feasibility conditions, we then minimize the expected power consumption of the D2D layer while maintaining the outage performance of both cellular layer and D2D layer. The motivation of minimizing the power consumption of the D2D layer is to ensuring quality-of-service or protecting the cellular layer from harmful interference caused by the D2D layer. Two power control scheme, upon transmit power is \emph{independent} or \emph{dependent} of channel conditions, are taken into account. We derive the feasibility conditions for the power-efficiency optimization problem for both independent and dependent power control cases. For the independent power control, we prove that fixed power control is optimal and we derive its closed-form. For the dependent power control case, we first show that the expected power consumption can be minimized if the power allocations are the deterministic functions of channel conditions. Then we propose an efficient method to find the close-to-optimal solution based on some approximations.

The rest of the paper are structured as follows. Section II describes the system model and problem formulation. Sections III and IV present our main results of independent and dependent power control schemes, respectively. Comprehensive numerical results are provided in Section V. Finally, Section VI concludes this paper.

\section{System Model}
\subsection{Network Model}

We consider the cellular network where the D2D users and cellular users coexist by the spectrum-sharing manner. Here we assume that the transmission mode, i.e., direct mode via D2D link and cellular mode via BS, is predetermined for each user. The locations of the D2D transmitters and cellular users are modeled as independent stationary Poisson Point Processes (PPP). Denote $\Phi_c $ ($\Phi_d$) and $\lambda_c $ ($\lambda_d$) as the locations and density of cellular (D2D) users, respectively. For simplicity, the distance between transmitter and receiver of each D2D (cellular) user is assumed to be fixed and denoted as $r_d$ ($r_c$).
%Similarly, the distance between each cellular user and its serving base station is assumed to be fixed and denoted as $r_c$.
In this paper, we consider that the D2D users reuse the uplink spectrum of the cellular users. The network is assumed to be interference limited, so that the background noise is neglected \cite{overlaid,SGLD}.

\subsection{Channel Model}

Denote $x_{kk'}$  as the distance between users $k$ and $k'$, $k\neq k'$.
The signal-to-interference ratio (SIR) at the receiver of D2D user $k$ is given by
\begin{equation}
SIR_k=\frac{P_kh_kr_d^{-\alpha}}{\sum_{k'\in\Phi_c}P_{k'}h_{kk'}|x_{kk'}|^{-\alpha}+\sum_{k'\in\Phi_d\setminus \{k\}}P_{k'}h_{kk'}|x_{kk'}|^{-\alpha}},
\end{equation}
where $\alpha$ is the path-loss exponent, $P_i$ is the transmit power of user $i$ and $h_{ij}$ denote the i.i.d Rayleigh fading coefficients with $\mathbb{E}[h_{ij}]=1$, where $\mathbb{E}[\cdot]$ is the expectation operator.
If $k$ is a cellular user, we just swap the subscripts $c$ by $d$.

The transmit powers are assumed to be independent random variables for each user. It is also assumed that the users in the same network have identical transmit power distribution, but the distribution may vary from one network to another.
We consider two scenarios. The first one is that the transmit power of each user is independent of channel conditions. We refer to this scenario as \emph{independent power control}. The second one is that the transmit power of each user is dependent on the channel condition between its transmitter and its receiver. This is referred to as \emph{dependent power control}.

According to properties of Palm distribution in \cite{SG}, the SIR distributions of all users in the same network are identical if the networks are stationary PPP and independent. Therefore, without loss of generality, in the following, our attention will focus on two \emph{typical} users, one for cellular user and the other for D2D user. The concept of typical is commonly used in the literature \cite{TC,TC2,OP}.
%Then for simplicity, we denote $SIR_c$ ($SIR_d$) as the SIR of typical cellular user's serving base station (typical D2D receiver).

We define the feasibility region $\mathcal{F}$ as the set of density pair $(\lambda_c,\lambda_d)$ such that make the outage probability is below specified thresholds $\epsilon_c$ and $\epsilon_d$ for the typical cellular and D2D users, respectively. Mathematically,
\begin{equation*}
\mathcal{F}(P_c,P_d)\triangleq\big\{(\lambda_c,\lambda_d):\Pr(SIR_c(P_c,P_d)\leq\theta_c)\leq\epsilon_c,\Pr(SIR_{d}(P_c,P_d)\leq\theta_d)\leq\epsilon_d\big\},
\end{equation*}
where $c$ and $d$ represent the typical cellular and D2D users, respectively; $\theta_c$ and $\theta_d$ is the SIR requirements for the typical cellular and D2D users, respectively.
Then the feasibility region can be expressed as:
\begin{equation}\label{feasibilityregion}
\mathcal{F}=\bigcup_{(P_c,P_d)}\mathcal{F}(P_c,P_d).
\end{equation}

\subsection{Problem Formulation}

Based on the Campbell's formula \cite{SG}, we know that if $A$ is any region with area 1, $\mathbb{E}[\sum_{k\in\Phi_d\bigcap A}P_k]=\lambda_d\mathbb{E}[P_d]$. That is to say $\lambda_d\mathbb{E}[P_d]$ is the average power consumption of D2D users per unit area. Hence, the goal of this paper
is to minimize the averaged power consumption of the typical D2D user while maintaining the
outage probabilities of the typical D2D and cellular users are below specified thresholds. The
problem can be formulated as
\begin{eqnarray}\label{minpower}
  \textbf{P1:}~~ && \min_{P_d}~\mathbb{E}[P_d] \\
  &&{\rm s.t.}~~ \Pr(SIR_i\leq\theta_i) \leq \epsilon_i, i\in\{c,d\}\label{constrainc}\\
  && ~~~~~~~0\leq P_ d \leq P_{d,max}.\label{constraine}
\end{eqnarray}

\section{Independent Case}

In this section, we analyze the independent case where the transmit power of the typical D2D
user is independent of channel conditions.

\subsection{Feasibility Region}

Due to the outage probability constraints, the problem \textbf{P1} is not always feasible. To make sure that \textbf{P1} will not have an empty set of solution, we should firstly analyze the feasibility region that depends on the system parameters.

\begin{lemma}\label{OPLA}
For the independent power control,
\begin{equation}\label{pr6}
\Pr(SIR_i\leq\theta_i)=1-\mathbb{E}\left[\exp\left\{-\frac{\phi_i(\lambda_c\mathbb{E}(P_c^{\delta})
+\lambda_d\mathbb{E}(P_d^{\delta}))}{P_i^\delta}\right\}\right],\footnote{$\lim_{x\to 0^+}\exp\{-1/x\}=0\triangleq\exp\{-1/0\}$.}\\
\end{equation}
where $\delta=2/\alpha$ and $\phi_i=\frac{\pi^2}{\sin(\pi\delta)}\delta\theta_i^\delta r_i^2$, $i\in\{c,d\}$.
\end{lemma}

\begin{IEEEproof}
  Please see Appendix \ref{COPIPC}.
\end{IEEEproof}

Note that \eqref{pr6} implies that the interference is infinity if $\delta\geq1$. This means that the independent power control cannot be used in the networks where the path-loss exponent $\alpha\leq2$. Thus we only consider the case $\delta<1$ in this section.

With the distribution of SIRs derived in \eqref{pr6}, we provide the feasibility region in the following theorem.

\begin{theorem}\label{PFR1t}
Assume that $\epsilon_d\leq1-\frac{1}{e}$, for given $P_c$, the feasibility region with respect to the independent power control is given by
\begin{eqnarray*}
\mathcal F=
\begin{cases}
\big\{(\lambda_c,\lambda_d):-\frac{\phi_d\lambda_d}{\ln(1-\epsilon_d)}+\frac{y_c\phi_c\lambda_c}{Q_c}\leq1\big\} & \mbox{if} \phantom{aaa} P_{d,max}>\bigg(\frac{Q_c\phi_d}{\phi_c\ln\frac{1}{1-\epsilon_d}}\bigg)^{\frac{1}{\delta}},\\
\big\{(\lambda_c,\lambda_d):\lambda_d+\lambda_c\frac{y_c}{P_{c,max}^\delta}\leq\frac{1}{\phi_d}\ln\frac{1}{1-\epsilon_d}\big\} & \mbox{if}\phantom{aaa} P_{d,max}\leq\bigg(\frac{Q_c\phi_d}{\phi_c\ln\frac{1}{1-\epsilon_d}}\bigg)^{\frac{1}{\delta}},
\end{cases}
\end{eqnarray*}
where $y_c=\mathbb{E}[P_c^\delta]$, $Q_c$ is given by lemma \ref{qd}.
\end{theorem}

\begin{IEEEproof}
  Please see Appendix \ref{PFR1proof}.
\end{IEEEproof}
The assumption of $\epsilon_i\leq1-\frac{1}{e}\approx0.63$ with $i\in\{c,d\}$ is reasonable since the target outage probabilities are not designed too large in practice.

The following corollary yields the way to achieve the feasibility region.
\begin{corollary}\label{PFR1tC}
Let $\epsilon_d\leq1-\frac{1}{e}$. For any pair $(\lambda_c,\lambda_d)$ that is in the feasibility region given by Theorem \ref{PFR1t}, the constraint \eqref{constrainc} holds if
\begin{eqnarray}
P_d=
\begin{cases}
\bigg(\frac{Q_c\phi_d}{\phi_c\ln\frac{1}{1-\epsilon_d}}\bigg)^{\frac{1}{\delta}}& \mbox{if} \phantom{aaa} P_{d,max}>\bigg(\frac{Q_c\phi_d}{\phi_c\ln\frac{1}{1-\epsilon_d}}\bigg)^{\frac{1}{\delta}},\label{RPC11}\\
P_{d,max}& \mbox{if}\phantom{aaa} P_{d,max}\leq\bigg(\frac{Q_c\phi_d}{\phi_c\ln\frac{1}{1-\epsilon_d}}\bigg)^{\frac{1}{\delta}}.\label{RPC12}
\end{cases}
\end{eqnarray}
\end{corollary}
The proof of Corollary \ref{PFR1tC} can be obtained from the proof of Theorem \ref{PFR1t}.

%\begin{IEEEproof}
%Recall that in the proof of Theorem \ref{PFR1t}, the feasibility region can be achieved when $P_d$ is a constant and $y_d=\min\{\frac{Q_c\phi_d}{\phi_c\ln\frac{1}{1-\epsilon_d}},P_{d,max}^\delta\}$.
%\end{IEEEproof}
%\emph{Remark:} Theorem \ref{PFR1t} indicates a requirement on networks to make the problem \textbf{P1} feasible. The corollary \ref{PFR1tC} indicates a specific example to support this feasibility.
%The proof of theorem \ref{PFR1} yields a corollary:
%\begin{corollary}
%If $P_c, \lambda_c$ has been known in priori and $P_c$ is independent of channel and network, then the maximal density of D2D users can be achieved when D2D users use fixed power at
%\end{corollary}
%\textcolor{red}{Best Power Scheme that maximize $\lambda_d$?}

\subsection{Minimizing Averaged Power Consumption}

In this subsection, we study the optimization problem \textbf{P1}. It is assumed that the density pair $(\lambda_c,\lambda_d)$ is given and feasible. $\epsilon_c,\epsilon_d\leq 1-\frac{1}{e}$, $\lambda_c\not=0$, and $\mathbb{E}[P_c^\delta]\not=0$. Thus, according to Lemma \ref{qd}, there exists a $Q_c$ and $R_c=\big(\frac{Q_c}{\phi_c}-\lambda_cy_c\big)\frac{1}{\phi_d}$, such that the problem becomes\footnote{Without loss of generality, it is also assumed that $P_{d,max}^\delta\geq R_c$.}
\begin{eqnarray}
  \textbf{P2:}~&&\min_{P_d}~~\mathbb{E}[P_d],\label{max3} \\
  &&~ {\rm s.t.}~~~\mathbb{E}[P_d^\delta]\leq R_c,\label{constrainc2}\\
&&~~~~~~~~\Pr(SIR_d\leq\theta_d)  \leq\epsilon_d,\label{constraind2}\\
&&~~~~~~~~0\leq P_ d  \leq P_{d,max}.\label{constraine2}
\end{eqnarray}

\begin{theorem}\label{lowestpower1}
The optimal solution for the problem \textbf{P2} is
\begin{equation}\label{pdo}
P_{d,opt}=P_{d0}=\left(\frac{\lambda_cy_c}{\frac{1}{\phi_d}\ln\frac{1}{1-\epsilon_d}-\lambda_d}\right)^{\frac{1}{\delta}} \phantom{aa} a.s.,
\end{equation}
\end{theorem}
where ``a.s." means with probability 1.

\begin{IEEEproof}
  Please see Appendix \ref{lowestpower1proof}.
\end{IEEEproof}

Note that the optimal power allocation in \eqref{pdo} is a constant, which means that the optimal power allocation is fixed and without any randomness of the networks.

It is remarkable that we assume $\epsilon_d\leq 1-\frac{1}{e}$ in above derivations. If without the assumption, our fixed power control will be relaxed to random power control as in \cite{RPC} and the feasibility region may be enlarged. However, $\epsilon_d\geq 1-\frac{1}{e}\approx0.63$ is not practical for outage threshold design.

\section{Dependent Case}

In this section, we analyze the dependent case where the transmit power of the typical D2D
user is dependent on the channel condition between its transmitter and receiver.

\subsection{Feasibility Region}
Like the previous section, we first study the feasibility region of the dependent power control case. However, the dependent case is more complex than the independent case. To make our analysis tractable, we adopt a lower bound of the expression of outage probability.
\begin{lemma}\label{OPLB}
For the dependent power control, a lower bound of outage probability is
\begin{equation}\label{lb}
\Pr(SIR_i\leq\theta)\geq1-\mathbb{E}[\exp\big\{-\psi_i(\lambda_c\mathbb{E}(P_c^\delta)+\lambda_d\mathbb{E}(P_d^\delta))(h_iP_i)^{-\delta}\big\}]
\end{equation}
where $\psi_i=\pi r_i^2\theta_i^\delta\Gamma(1+\delta),i\in\{c,d\}$, $\Gamma(x)=\int_0^\infty t^{x-1}e^{-t}dt$ and $h_i$ is the channel gain between the transmitter and receiver, $i\in\{c,d\}$.
\end{lemma}

\begin{IEEEproof}
  Please see Appendix \ref{COPDPC}.
\end{IEEEproof}

The tightness of this lower bound can be proved using the similar method in \cite{TC2} and the details are omitted here.
It is observed that if $\epsilon_i$ is small, $\theta_i$ is also small. According to Jensen's inequality and the fact that function $e^{-x}$ is close to be a linear function, we further introduce an approximation for \eqref{lb} to ease the analysis.  %Meanwhile, we further denote $\psi_i=\pi r_i^2\theta^\delta\mathbb{E}(h^\delta),i\in\{c,d\}$.
\begin{eqnarray}
\Pr(SIR_i\leq\theta)&\geq&1-\mathbb{E}\big[\exp\big\{-\psi_i(\lambda_c\mathbb{E}(P_c^\delta)+\lambda_d\mathbb{E}(P_d^\delta))(h_iP_i)^{-\delta}\big\}\big],\nonumber\\
&\approx& 1-\exp\big\{-\psi_i(\lambda_c\mathbb{E}(P_c^\delta)+\lambda_d\mathbb{E}(P_d^\delta))\mathbb{E}(h_i^{-\delta}P_i^{-\delta})\big\},\label{approx}\\
&\triangleq&\underline{Pr_i},\nonumber
\end{eqnarray}
where $i\in\{c,d\}$. Notice that the approximation is also similarly used in \cite{FPC} where a specific power control scheme in a single network.

With the approximation \eqref{approx}, the outage probability constraint becomes
\begin{equation}\label{approxconstrain}
\underline{Pr_i}\leq\epsilon_i,i\in\{c,d\},
\end{equation}
and the feasibility region can be characterized by the following theorem.
\begin{theorem}\label{PFR2t}
The feasibility region with the dependent power control is given by
\begin{equation}
\mathcal F=\left\{(\lambda_c,\lambda_d):\lambda_cy_c+\lambda_d\frac{\psi_d\Gamma(1-\frac{\delta}{2})^2
\ln(1-\epsilon_c)}{\psi_c\mathbb{E}[P_c^{-\delta}h_c^{-\delta}]\ln(1-\epsilon_d)}\leq\psi_c^{-1}
\mathbb{E}[P_c^{-\delta}h_c^{-\delta}]^{-1}\ln\frac{1}{1-\epsilon_c}\right\}.
\end{equation}
\end{theorem}

\begin{IEEEproof}
  Please see Appendix \ref{PFR2tproof}.
\end{IEEEproof}

\begin{corollary}\label{PFR2tC}
For any density $(\lambda_c,\lambda_d)$ that is in the feasibility region given by Theorem \ref{PFR2t}, \eqref{approxconstrain} holds if
\begin{eqnarray*}
P_d=\left[\frac{\psi_d\Gamma{(1-\frac{\delta}{2}})\ln(1-\epsilon_c)}{\psi_c\mathbb{E}
[P_c^{-\delta}h_c^{-\delta}]\ln(1-\epsilon_d)}\right]^{\frac{1}{\delta}}h_d^{-\frac{1}{2}}.
\end{eqnarray*}
\end{corollary}

Corollary \ref{PFR2tC} can be regarded as fractional power control with the exponent $0.5$. The properties of fractional power control has been discussed in \cite{FPC},  where the authors  prove that the fractional power control with exponent $0.5$ can minimize the approximation \eqref{approx} among all the fractional power control policies. Our result in Corollary \ref{PFR2tC} demonstrates another optimality: the fractional power control with exponent $0.5$ also leads to the largest feasibility region among all the dependent power control policies.
%\emph{Remark:} Theorem \ref{PFR2t} indicates a requirement on networks to make the problem \textbf{P1} feasible. The corollary \ref{PFR1tC} indicates a specific example to support this feasibility.
%\begin{IEEEproof}
%Recall that in the proof of Theorem \ref{PFR2}, the feasibility region can be achieved when $P_d$ is given by Lemma \ref{min} and $y_d=\frac{\psi_d\Gamma{(1-\frac{\delta}{2}})\ln(1-\epsilon_c)}{\psi_c\mathbb{E}(P_c^{-\delta}h_c^{-\delta})\ln(1-\epsilon_d)}$.
%\end{IEEEproof}
%\emph{Remark:}In the assumption of this theorem, we only assume the independence of other channel and network. The transmit power can be dependent on other factors while the results also holds.

\subsection{Minimizing Averaged Power Consumption}
In this subsection, we consider the optimization problem \textbf{P1} under dependent power control. Like \cite{FPC}, we drop the peak power constrain. We also assume that $(\lambda_c,\lambda_d)$ is given and feasible. The problem can be rewritten as
\begin{eqnarray}
\textbf{P3:}~&&\min_{P_d}\mathbb{E}[P_d],\label{max5}\\
  &&~ {\rm s.t.}~~\mathbb{E}[P_d^\delta]\leq \bigg(-\frac{\ln(1-\epsilon_c)}{\mathbb{E}[P_c^{-\delta}h_c^{-\delta}]\psi_c}-\lambda_cy_c\bigg)\frac{1}{\lambda_d},\label{constrainc5}\\
&&~~~~~~~~(\lambda_cy_c+\lambda_d\mathbb{E}[P_d^\delta])\mathbb{E}[P_d^{-\delta}h_d^{-\delta}]  \leq\frac{1}{\psi_d}\ln\frac{1}{1-\epsilon_d},\label{constraind5}\\
&&~~~~~~~~0\leq P_ d.\label{constraine5}
\end{eqnarray}

%\begin{lemma}
%Consider a new constrain:
%\begin{equation}
%(\lambda_cy_c+\lambda_d\mathbb{E}{P_d^\delta})\mathbb{E}(P_d^{-\delta}h^{-\delta})=\ln\frac{1}{1-\epsilon_d}\label{constraind6}\\
%\end{equation}
%If $P^\star$ satisfy \eqref{constrainc5}, \eqref{constraind5} and \eqref{constraind5}, then there exists $P^{\star\star}$, such that
%\eqref{constrainc5}, \eqref{constraind6} and \eqref{constraine5} hold. Meanwhile, $\mathbb{E}(P^\star)\geq\mathbb{E}(P^{\star\star})$
%\end{lemma}
%\begin{IEEEproof}
%If $P^\star$ satisfy \eqref{constraind6}, then set $P^{\star\star}=P^\star$. Otherwise we need to construct a new $P^{\star\star}$.
%Consider the function $f(t)=(\lambda_cy_c+\lambda_d\mathbb{E}{(tP_d)^\delta})\mathbb{E}((tP_d)^{-\delta}h^{-\delta})=(\frac{\lambda_cy_c}{t^{\delta}}+\lambda_d\mathbb{E}{(P_d)^\delta})\mathbb{E}((P_d)^{-\delta}h^{-\delta})$.
%$f(t)$ is a continuous decreasing function and $\lim_{t\to0f(t)=\infty$. Then there exists $t_0$ such that $f(t_0)=\ln\frac{1}{1-\epsilon_d}$
%\end{IEEEproof}
\begin{lemma}\label{function}
If the optimal power allocation $P^\star$ satisfies \eqref{constrainc5}-\eqref{constraine5}, then there exists $P^{\star\star}=P^{\star\star}(h)$ that is a deterministic function, such that \eqref{constrainc5}-\eqref{constraine5} hold. Meanwhile, $\mathbb{E}[P^\star]\geq\mathbb{E}[P^{\star\star}]$.
\end{lemma}

\begin{IEEEproof}
  Please see Appendix \ref{functionproof}.
\end{IEEEproof}

The above Lemma reveals that we should only focus on the deterministic functions, because the additional randomness will not improve performance.

Due to extremely sophisticated structure of the function spaces, finding the extremal point for functionals is very nontrivial. We then propose an efficient method to solve the power control problem in a suboptimal way by conducting three ordinary relaxations. Firstly, instead of considering the function $P_d(h)$ in $[0,\infty)$, we consider it in $[0,M]$ where $M$ is large. This is possible since channel gain can not be infinity in practice. For the points outside this interval, the function is defined to be $0$. This relaxation is plausible because when the $M$ is large, we can cover almost all the status of channel gain. The second relaxation is to consider the function to be piecewise constant with each interval of constant very small. Based on these two relaxation, the function is converted to be:
\begin{equation*}
P_d(h)=\sum_{i=0}^{N-1}p_i\mathbbm{1}_{[x_i,x_{i+1})}(h),
\end{equation*}
where $0=x_0<x_1<x_2...<x_N=M$. In this case the choices of $\{x_i\}_{i=0}^N$ decide the accuracy of the relaxation.
Hence, \textbf{P3} becomes:
\begin{eqnarray}
&& \min_p\sum_{i=0}^{N-1}a_ip_i, \\
&&{\rm s.t.}~~ (\lambda_cy_c+\lambda_d\sum_{i=0}^{N-1}a_ip_i^\delta)(\sum_{i=0}^{N-1}c_ip_i^{-\delta})\leq\frac{1}{\psi_d}\ln\frac{1}{1-\epsilon_d},\\
&& ~~~~~~~\sum_{i=0}^{N-1}a_ip_i^\delta\leq \bigg(-\frac{\ln(1-\epsilon_c)}{\mathbb{E}[P_c^{-\delta}h_c^{-\delta}]\psi_c}-\lambda_cy_c\bigg)\frac{1}{\lambda_d},\\
&& ~~~~~~~0\leq P_ d.
\end{eqnarray}
where $a_i=\int_{x_i}^{x_{i+1}}e^{-h}dh$, $c_i=\int_{x_i}^{x_{i+1}}h^{-\delta}e^{-h}dh$. This is a geometry programming problem and can be solved efficiently since geometry programming problems can be transformed into convex problems.
\section{Numerical Results}
%bound tightness
In this section, the implications of independent and dependent power control are illustrated through plots and figures. As default parameters, the simulation assumes:
\begin{equation*}
\theta_c=\theta_d=0.1, \epsilon_c=\epsilon_d=0.01, r_c=r_d=1, \alpha=2/0.75\approx2.67
\end{equation*}
\subsection{Feasibility Region under Different Schemes}

\begin{figure}
\center
\includegraphics[scale=0.65]{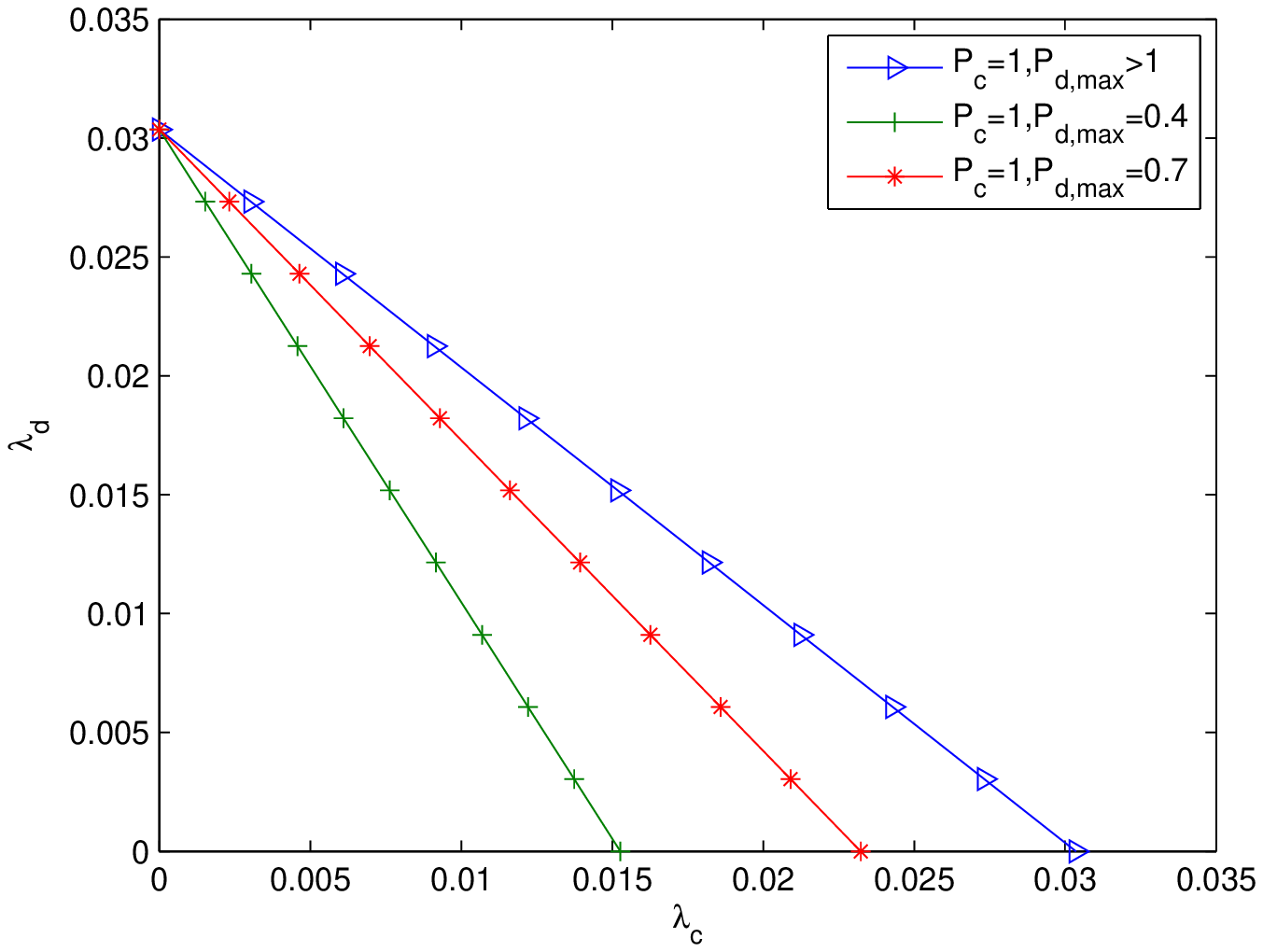}
\center\caption{Feasibility region when $P_c$ is given.}\label{PFR1}
\end{figure}

If $P_c$ is given, Theorems \ref{PFR1t} and \ref{PFR2t} describe the feasibility region under independent and dependent control respectively.  Fig. \ref{PFR1} reveals the feasibility versus peak power constraint for the independent control. It can be observed that a) the peak power constraint of $P_d$ can significantly reduce the supported $\lambda_c$; b) when the peak power constraint exceeds a certain level, it will not have any effect on the feasibility region.

\begin{figure}
\center
\includegraphics[scale=0.65]{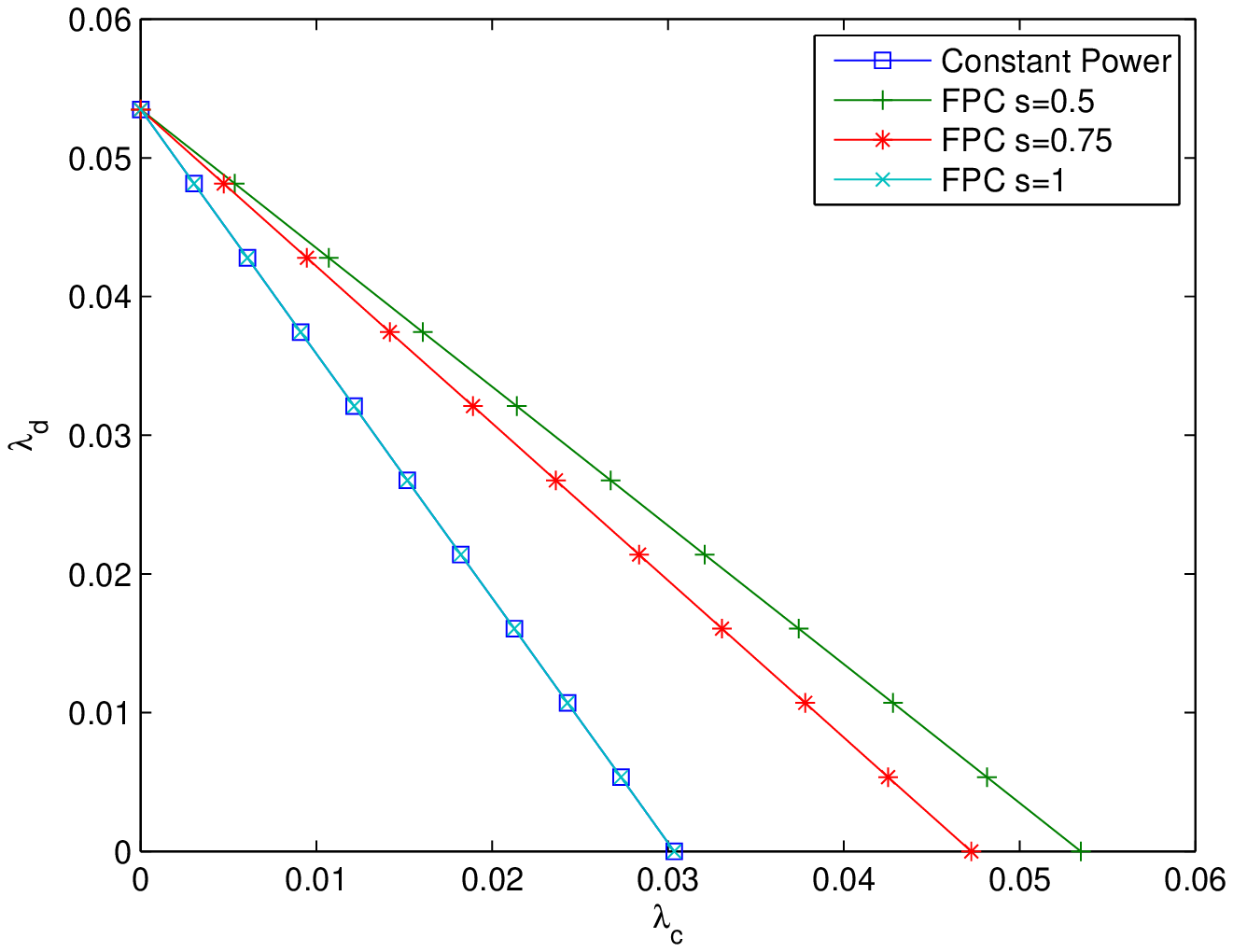}
\center\caption{Feasibility region when $P_c$ is given.}\label{PFR2}
\end{figure}

On the other hand, Fig. \ref{PFR2} illustrates the case of dependent power control when different $P_c$'s are given. Recall Theorem \ref{PFR2} that if then the feasibility region is calculated for a specific $P_c$, any scalar of $P_c$ will not change this region. Thus, we omit the coefficient of $P_c$. We study the feasibility region when $P_c=h^{-s}$, i.e., $P_c$ is using fractional power control \cite{FPC}. We can observe from the figure: a) channel inverse and constant power control have the same feasibility region, b) $s=\frac{1}{2}$ outperforms others that was considered.
%In this section, we set $\theta=0.1,\epsilon_c=\epsilon_d=0.01,r_c=r_d=1$
%We first consider the case when both $P_c$ and $P_D $ have not been designed in priori.
%
%From Fig. \ref{FR1vsFR2}, we can see that the boundaries under the same coefficients between different fading exponents and different schemes are parallel lines. Besides, the dependent power control yields bigger feasibility region. According to the remarks of Theorem \ref{FR2}, the feasibility region of dependent power control is larger than independent power control in a factor of $\frac{\Gamma(1-\frac{2}{\alpha})}{\Gamma(1-\frac{1}{\alpha})^2}$. This function is demonstrated in Fig. \ref{ratio}. Fig. \ref{ratio} shows that a) the ratio is always greater than $1$, b) when $\alpha$ is approaching to 2, the ratio is taking a boost, c) when $\alpha$ is far away from 2, dependent power control will not bring much performance gain. The boosting is mainly because under dependent power control $\alpha\leq2$ is supported. Under independent power control, when $\alpha\to2$ the feasibility region will tend to be empty.
%\begin{figure}
%\center
%\includegraphics[scale=0.65]{FR1vsFR2.eps}
%\caption{Feasibility Region under Different Power Control Scheme}\label{FR1vsFR2}
%\end{figure}
%\begin{figure}
%\center
%\includegraphics[scale=0.65]{ratio.eps}
%\center\caption{Scaling Factor, the line in the bottom is indicating where 1 is}\label{ratio}
%\end{figure}

%We then demonstrate how the feasibility region will change when only one of $P_c$ and $P_d$ is determined in priori with further setting $\alpha=2.67$.


\subsection{Numerical Results of Minimal Averaged Power}
In this subsection, we consider two problems.

\subsubsection{Convergence Problem}\label{convergence}
\begin{figure}
\center
\includegraphics[scale=0.65]{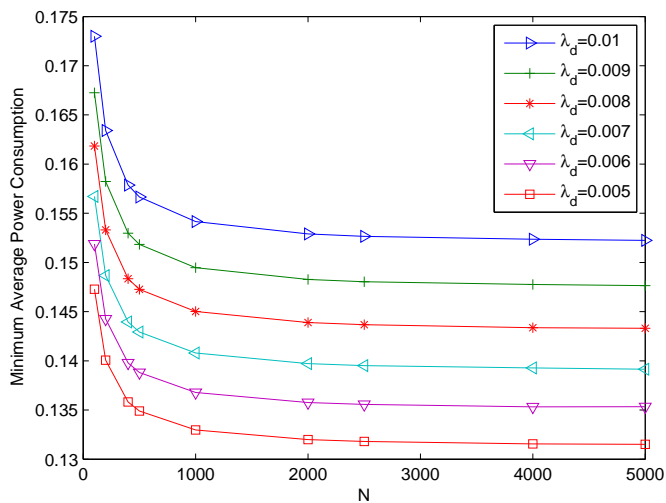}
\center\caption{Convergence with respect to $N$.}\label{convergencefig}
\end{figure}

First, we consider the minimal averaged power consumption versus $N$. Note that $h^{-\delta}$ changes really fast when $h$ is close to zero. Thus, we design $\{x_i\}_{i=0}^N$ as follows:
\begin{eqnarray*}
x_n=
\begin{cases}
0.002 \frac{n}{N}, & \text{if} \phantom{aaa} n\leq\frac{N}{2},\\
0.001+(M-0.001) \frac{n}{N}, & \text{if} \phantom{aaa} n\geq\frac{N}{2}+1.
\end{cases}
\end{eqnarray*}
we will show that if $N$ is sufficient large, the minimal averaged power consumption will finally converge. The result is given in Fig. \ref{convergencefig}. $P_c=1$ is set and we simulate for different $\lambda_d$'s. From the figure, it is observed that when $N$ is greater than $2500$, the minimal average power consumption will converge. This shows the effectiveness of our proposed method in Section IV-B.

\begin{figure}
\center
\includegraphics[scale=0.65]{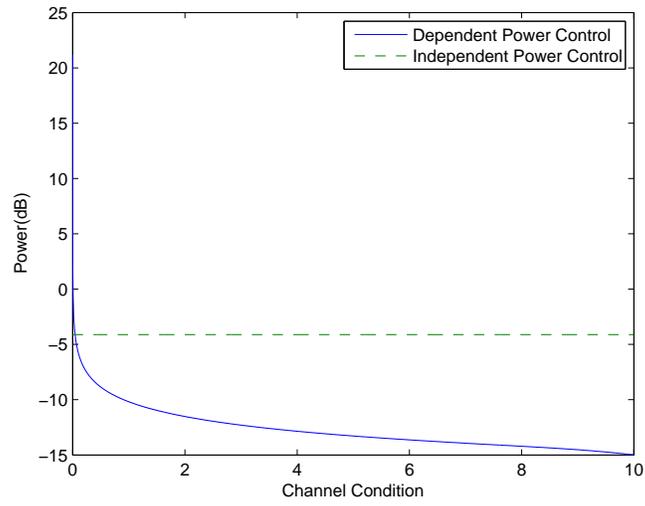}
\center\caption{Independent power control and dependent power control for $\lambda_d=0.01$.}\label{pcschemes}
\end{figure}

\begin{figure}
\center
\includegraphics[scale=0.65]{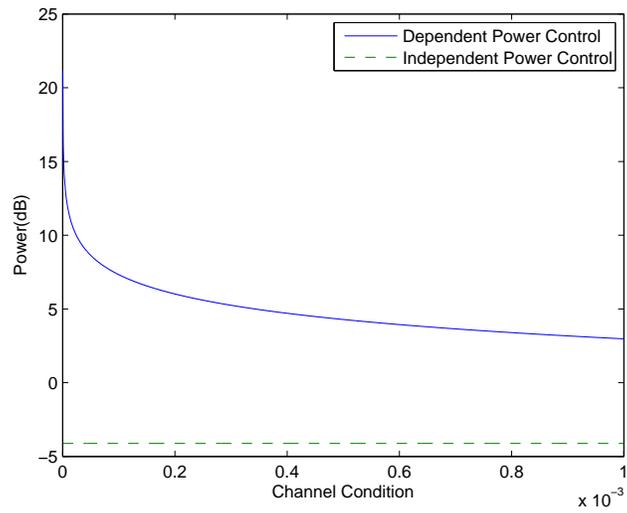}
\center\caption{Independent power control and dependent power control for $\lambda_d=0.01$, when channel gain is small.}\label{pcschemessmallh}
\end{figure}

Fig. \ref{pcschemes} shows the performance of power control schemes when $N=5000$ and $\lambda_d=0.01$. It can be observed that for most of the channel status, the power under independent power control is consumed more than the dependent case. However, when the channel gain is very close to zero, the power under dependent power control increases very fast. Fig. \ref{pcschemessmallh} demonstrates this increasing more accurately.

\subsubsection{Comparison between Independent and Dependent Power Control}

\begin{figure}
\center
\includegraphics[scale=0.65]{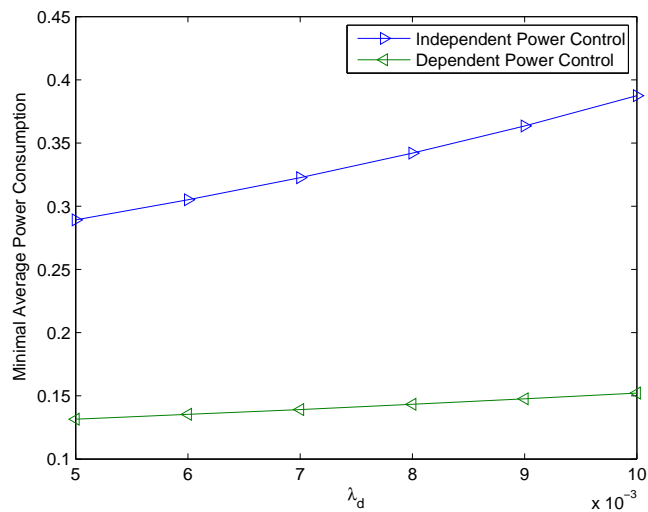}
\center\caption{Feasibility region when $P_c$ is designed in priori.}\label{powercompare}
\end{figure}

In this part, we compare the performance of the independent and dependent power control policies. Intuitively, dependent power control should have better performance. We investigate the relationship between $\lambda_d$ and the minimal power consumption. This comparison is conducted under the condition that $P_c=1$ is a constant. Here $N$ is set to be $5000$. The result is shown in Fig. \ref{powercompare}. From the figure, it is clear that dependent power control saves about 50\% power than independent power control. However, there is no free lunch. In order to save this energy, the D2D users should know the channel status at the transmitter side.
%\subsubsection{the Effect of the Fading Exponent $\alpha$}

\section{Conclusion}

In this paper, we considered the power minimization problem of D2D users to guarantee outage probabilities of both D2D and cellular users. For the random networks, two power control schemes, namely independent and dependent power control, were proposed based on stochastic geometry. For these two schemes, we first analyzed the feasibility of the power-efficiency problem. Then optimal and close-to-optimal solutions for the two schemes were proposed respectively. Numerical results showed that the dependent power control saves about 50\% power than the independent power control.

% or


\appendices
\section{Calculation of Outage Probability under Independent Power Control}\label{COPIPC}
\begin{IEEEproof}
If $k$ is a typical D2D user, then
\begin{eqnarray*}
\Pr(SIR_k>\theta_d) & \overset{\text{(a)}}{=} & \Pr\big(\frac{h_kP_dr_d^{-\alpha}}{\sum_{k'\in\Phi_c}P_ch_{kk'}|x_{kk'}|^{-\alpha}+\sum_{k'\in\Phi_d-\{k\}}P_dh_{kk'}|x_{kk'}|^{-\alpha}}>\theta_d\big),\\
&=&
\Pr(h_k>\frac{\theta_d r_d^\alpha}{P_d}\big(\sum_{k'\in\Phi_c}P_ch_{kk'}|x_{kk'}|^{-\alpha}+\sum_{k'\in\Phi_d-\{k\}}P_dh_{kk'}|x_{kk'}|^{-\alpha}\big)),\\
&\triangleq&\mathbb{E}_{P_d}E_{\Phi_c}E_{\Phi_d}\exp\{-\frac{\theta_d r_d^\alpha}{P_d}(I_1+I_2)\},\\
& \overset{\text{(b)}}{=} & \mathbb{E}_{P_d}\big\{\mathbb{E}_{\Phi_c}\exp\{-\frac{\theta_d r_d^\alpha}{P_d}I_1\}\mathbb{E}_{\Phi_d}\exp\{-\frac{\theta_d r_d^\alpha}{P_d}I_2\}\big\},\\
& \overset{\text{(c)}}{=} & \mathbb{E}_{P_d}\mathcal{L}_{I_1}(s)\big|_{s=\frac{\theta_d r_d^\alpha}{P_d}}\mathcal{L}_{I_2}(s)\big|_{s=\frac{\theta_d r_d^\alpha}{P_d}},\\
& \overset{\text{(d)}}{=} & \mathbb{E}\exp\big\{-\frac{\phi_d(\lambda_c\mathbb{E}(P_c^{\delta})+\lambda_d\mathbb{E}(P_d^{\delta}))}{P_d^\delta}\big\}.
\end{eqnarray*}
where $\phi_d=\frac{\pi^2}{\sin(\pi\delta)}\delta\theta_d^\delta r_d^2,\delta=\frac{2}{\alpha}$,
(a) from the fact that power is independent of channel,
(b) from the fact that power, $\Phi_c$, $\Phi_d$ are independent,
(c) from the definition of Laplace transform,
(d) from \cite{RPC2} Lemma 1. The other part of this lemma can be deduced similarly.
\end{IEEEproof}
\section{Calculation of Outage Probability under Dependent Power Control}\label{COPDPC}
\begin{IEEEproof}
We extend the proof in \cite{OP}, where only one network is considered.
If $k$ is the typical D2D user,
then
\begin{eqnarray*}
SIR_k&=&\frac{h_kP_kr_d^{-\alpha}}{\sum_{k'\in\Phi_c}h_{kk'}P_{k'}|x_{kk'}|^{-\alpha}+\sum_{k'\in\Phi_d-\{k\}}h_{kk'}P_{k'}|x_{kk'}|^{-\alpha}}.
\end{eqnarray*}
Set
\begin{eqnarray*}
\omega&=&h_kP_kr_d^{-\alpha},\\
\Phi_{\theta_d,\omega,c}&=&\big\{k'\in\Phi_c:\frac{1}{\omega}h_{kk'}P_{k'}|x_{kk'}|^{-\alpha}\geq\frac{1}{\theta_d}\big\},\\
\Phi_{\theta_d,\omega,d}&=&\big\{k'\in\Phi_d:\frac{1}{\omega}h_{kk'}P_{k'}|x_{kk'}|^{-\alpha}\geq\frac{1}{\theta_d}\big\},\\
Y_{\theta_d,\omega}&=&\frac{1}{\omega}\bigg(\sum_{k'\in\Phi_{\theta_d,\omega,c}}h_{kk'}P_{k'}|x_{kk'}|^{-\alpha}+\sum_{k'\in\Phi_{\theta_d,\omega,d}}h_{kk'}P_{k'}|x_{kk'}|^{-\alpha}\bigg).
\end{eqnarray*}
Therefore,
\begin{eqnarray*}
\Pr(SIR_k\leq\theta_d)&\geq&\Pr(Y_{\theta_d,\omega}\geq\frac{1}{\theta_d}),\\
&=&1-\Pr(\Phi_{\theta_d,\omega,c}=\varnothing,\Phi_{\theta_d,\omega,d}=\varnothing),\\
& {=} & 1-\mathbb{E}_{\omega}\Pr(\Phi_{\theta_d,\omega,c}=\varnothing,\Phi_{\theta_d,\omega,d}=\varnothing|\omega),\\
& \overset{\text{(a)}}{=} & 1-\mathbb{E}_{\omega}\big(\Pr(\Phi_{\theta_d,\omega,c}=\varnothing|\omega)\Pr(\Phi_{\theta_d,\omega,d}=\varnothing|\omega)\big),\\
& \overset{\text{(b)}}{=} & 1-\mathbb{E}_{\omega}\big(\exp\big\{-\int_{\mathbb{R}^2}\mu_{\theta_d,\omega,c}(x)dx\big\}\exp\big\{-\int_{\mathbb{R}^2}\mu_{\theta_d,\omega,d}(x)dx\big\}\big),\\
& \overset{\text{(c)}}{=} &
1-\mathbb{E}\big(\exp\big\{-(\lambda_c\mathbb{E}(h_c^\delta P_c^\delta)+\lambda_d\mathbb{E}(h_d^\delta P_d^\delta))\pi(h_kP_k)^{-\delta}r_d^2\theta_d^\delta\big\}\big),\\
& \overset{\text{(d)}}{=} &
1-\mathbb{E}\big(\exp\big\{-(\lambda_c\mathbb{E}(P_c^\delta)+\lambda_d\mathbb{E}(P_d^\delta))\pi\mathbb{E}(h^\delta)(h_kP_k)^{-\delta}r_d^2\theta_d^\delta\big\}\big),
\end{eqnarray*}
where (a) from $\omega,\Phi_c,\Phi_d$ are independent, (b) follows from (103) in \cite{OP}, (c) follows from (104),(105) in \cite{OP}, (d) from the fact that power is a random variable which is independent of the channel that is not between the transmitter receiver pair and $\delta=\frac{2}{\alpha}$. If $k$ is a typical D2D cellular, the proof is identical.
\end{IEEEproof}
\section{Proof of Theorem \ref{PFR1t}}\label{PFR1proof}
Before proving theorem \ref{PFR1t}, we need some lemmas.
\begin{lemma}\label{continuity}
The function,
\begin{equation*}
f(x)\triangleq\mathbb{E}\exp\{-\frac{x}{\xi}\},
\end{equation*}
is a continuous function of $x\geq0$ for an arbitrary random variable $\xi$ with $\xi\geq0$.
\end{lemma}
\begin{IEEEproof}
%Set $c,c_0\geq0$ then:
%\begin{eqnarray*}
%&&\lim_{c\to c_0}\big|\mathbb{E}\exp\{-\frac{c}{\xi}\}-\mathbb{E}\exp\{-\frac{c_0}{\xi}\}\big|\\
%& \overset{\text{(a)}}{=} & \lim_{c\to c_0}\big|\int^\infty_0(1-F(\frac{c}{x}))e^{-x}dx-\int^\infty_0(1-F(\frac{c_0}{x}))e^{-x}dx\big|\\
%& \overset{\text{(b)}}{=} & \lim_{c\to c_0}\big|c\int^\infty_0(1-F(\frac{1}{y}))e^{-cy}dy-c_0\int^\infty_0(1-F(\frac{1}{y}))e^{-c_0y}dy\big|\\
%&=& \lim_{c\to c_0}\big|(c-c_0)\int^\infty_0(1-F(\frac{1}{y}))e^{-cy}dy-c_0\int^\infty_0(1-F(\frac{1}{y}))(e^{-c_0y}-e^{-cy})dy\big|\\
%&\leq& \lim_{c\to c_0}\big|(c-c_0)\int^\infty_0(1-F(\frac{1}{y}))e^{-cy}dy\big|+\lim_{c\to c_0}\big|c_0\int^\infty_0(1-F(\frac{1}{y}))(e^{-c_0y}-e^{-cy})dy\big|\\
%&\leq&\lim_{c\to c_0}c_0\int^\infty_0(1-F(\frac{1}{y}))\big|(e^{-c_0y}-e^{-cy})\big|dy\\
%& \overset{\text{(c)}}{=} & c_0\int^\infty_0(1-F(\frac{1}{y}))\lim_{c\to c_0}\big|(e^{-c_0y}-e^{-cy})\big|dy=0
%\end{eqnarray*}
%where (a) see the proof of Lemma \ref{max}, (b) set $y=cx$, $y=c_0x$ separately, (c) dominated convergence theorem \cite{probability}.
Note that $0\leq\exp\{-\frac{x}{\xi}\}\leq 1$, then according to dominated convergence theorem \cite{probability}, we have, for $x,x_0\geq0$, $\lim_{x\to x_0}\mathbb{E}\exp\{-\frac{x}{\xi}\}=\mathbb{E}\lim_{x\to x_0}\exp\{-\frac{x}{\xi}\}=\mathbb{E}\exp\{-\frac{x_0}{\xi}\}$.
\end{IEEEproof}
\begin{lemma}\label{qd}
If $P_c$ has been designed in priori and is independent of the networks as well as channel fading, $\phi_c(\lambda_c\mathbb{E}(P_c^{\delta})+\lambda_d\mathbb{E}(P_d^{\delta}))\geq0$, then there exists a constant $Q_c$ such that $\Pr(SIR_c>\theta)>1-\epsilon_c$ if and only if:
\begin{equation*}
0\leq\phi_c(\lambda_c\mathbb{E}(P_c^{\delta})+\lambda_d\mathbb{E}(P_d^{\delta}))<Q_c.
\end{equation*}
\end{lemma}
\begin{IEEEproof}
Set $q=\phi_c(\lambda_c\mathbb{E}(P_c^{\delta})+\lambda_d\mathbb{E}(P_d^{\delta}))$, $\Pr(SIR_c>\theta)=\mathbb{E}\exp\{-\frac{q}{P_c^\delta}\}\triangleq f(q)$. According to Lemma \ref{continuity}, $f(q)$ is a continuous function with respect to $q$. We then consider the inequality $f(q)>1-\epsilon_c$.

Since $(1-\epsilon_c,\infty)$ is an open set, then continuity implies that $\{q:f(q)>1-\epsilon_c\}$ is an open set. On the other hand, dominated convergence theorem ensures that $\lim_{q\to \infty}\mathbb{E}\exp\{-\frac{q}{P_c^\delta}\}=0$. So there exists a positive number $Q_0<\infty$ such that
\begin{equation*}
Q_0=\sup\{q:f(q)>1-\epsilon_c\}.
\end{equation*}

That is to say, if $f(q)>1-\epsilon_c$, then $0\leq q<Q_0$.
Meanwhile, it is easy to show that $f(q)$ is non-increasing with $q$. Thus for all $q,0\leq q<Q_0$ we have
\begin{equation*}
f(q)>1-\epsilon_c
\end{equation*}

Therefore, the $Q_0$ is the $Q_c$ we are looking for and this lemma is proved.
\end{IEEEproof}
\emph{Remark:} In Lemma \ref{qd}, if $\Pr(SIR_c>\theta)>1-\epsilon_c$ is changed into $\Pr(SIR_c>\theta)\geq1-\epsilon_c$, then $0\leq\phi_c(\lambda_c\mathbb{E}(P_c^{\delta})+\lambda_d\mathbb{E}(P_d^{\delta}))<Q_c$ is changed into $0\leq\phi_c(\lambda_c\mathbb{E}(P_c^{\delta})+\lambda_d\mathbb{E}(P_d^{\delta}))\leq Q_c$ and the proof is similar.
As for $Q_c$, because of monotonicity, the $Q_c$ can be found through numerical method such as bisection method \cite{bisection}.

\begin{lemma}\label{max}
If $\xi\geq0$ is a random variable, $c\geq0$ and $\xi_{max}\geq1$ are constants, consider this optimization problem:
\begin{eqnarray*}
  && \max_\xi~ \mathbb{E}[\exp\{-\frac{c}{\xi}\}] \\
  &&{\rm s.t.}~~~ \mathbb{E}\xi=1\\
  && ~~~~~~~\xi\leq\xi_{max}.
\end{eqnarray*}
The solution of this optimization problem follows:
If $F(x)$ is the cumulative distribution function(CDF) of the optimal $\xi$ then,\\
\begin{eqnarray*}
F(x)=
\begin{cases}
\mathbbm{1}_{[1,\infty)}(x) & \text{if} \phantom{aaa} c\leq 1,\\
(1-\frac{1}{c})\mathbbm{1}_{[0,c)}(x)+\mathbbm{1}_{[c,\infty)}(x) & \text{if} \phantom{aaa} 1<c\leq\xi_{max},\\
(1-\frac{1}{\xi_{max}})\mathbbm{1}_{[0,\xi_{max})}(x)+\mathbbm{1}_{[\xi_{max},\infty)}(x) & \text{if} \phantom{aaa} c>\xi_{max}.
\end{cases}
\end{eqnarray*}
\end{lemma}
\begin{IEEEproof}
Set $\eta$ is a random variable with probability density function(PDF) $f_{\eta}(x)=e^{-x}$, and $\eta$ is independent of $\xi$.
Then we have:
\begin{equation*}
\Pr(\eta\xi>c)=\Pr(\eta>\frac{c}{\xi})=\mathbb{E}\exp\{-\frac{c}{\xi}\}.
\end{equation*}
On the other hand,
\begin{equation*}
\Pr(\eta\xi>c)=\Pr(\xi>\frac{c}{\eta})=\int^\infty_0(1-F(\frac{c}{x}))e^{-x}dx.
\end{equation*}
Then the result comes directly from Theorem 1 in \cite{RPC3}
\end{IEEEproof}
%Similarly, it can be deduced from Corollary 1 in \cite{RPC3}:
%\begin{lemma}\label{max2}
%if $\xi\geq0$ is a random variable and $c\geq0$ is a constant, consider this optimization problem:
%\begin{equation*}
%\max_\xi \mathbb{E}\exp\{-\frac{c}{\xi}\}
%\end{equation*}
%subject to:
%\begin{eqnarray*}
%\mathbb{E}\xi=1
%\end{eqnarray*}
%The solution of this optimization problem follows:
%If $F(x)$ is the CDF of the optimal $\xi$ then,\\
%\begin{eqnarray*}
%F(x)=
%\begin{cases}
%\mathbbm{1}_{[1,\infty)}(x) & \text{if} \phantom{aaa} c\leq 1,\\
%(1-\frac{1}{c})\mathbbm{1}_{[0,c)}(x)+\mathbbm{1}_{[c,\infty)}(x) & \text{if} \phantom{aaa} c>1.\\
%\end{cases}
%\end{eqnarray*}
%\end{lemma}

Then we can start to proof theorem \ref{PFR1t}
\begin{IEEEproof}
Set $P_{d,y_d}=\max_{\mathbb{E}[P_d^\delta]=y_d}\Pr(SIR_d>\theta),y_c=\mathbb{E}[P_c^\delta]$, according to \eqref{feasibilityregion} we have:
\begin{eqnarray*}
\mathcal{F}&=&\big\{(\lambda_c,\lambda_d)\succeq0:\mbox{$\exists$ random varible $P_d$ such that \eqref{constrainc} hold}\big\},\\
&{=} & \big\{(\lambda_c,\lambda_d)\succeq0:\mbox{$\exists$ random varible $P_d$ such that $\Pr(SIR_d\leq\theta_d) \leq \epsilon_d$ and} \\ &&\mbox{$0\leq\phi_c(\lambda_c\mathbb{E}[P_c^{\delta}]+\lambda_d\mathbb{E}[P_d^{\delta}])\leq Q_c$ hold}\big\},\\
&=&\bigcup_{y_d}\big\{(\lambda_c,\lambda_d)\succeq0:\mbox{$\exists$ random varible $P_d$ with $\mathbb{E}[P_d^\delta]=y_d$ such that} \\&&\Pr(SIR_d(y_c,y_d)\leq\theta_d) \leq \epsilon_d \mbox{and $0\leq\phi_c(\lambda_cy_c+\lambda_dy_d)\leq Q_c$ hold}\big\},\\
&=&\bigcup_{y_d}\big\{(\lambda_c,\lambda_d)\succeq0:P_{d,y_d}\geq1-\epsilon_d,0\leq\phi_c(\lambda_cy_c+\lambda_dy_d)\leq Q_c\big\}.
\end{eqnarray*}

We calculate $P_{d,y_d}$. Set
\begin{equation*}
q=\frac{P_d^\delta}{y_d},s=\frac{\phi(\lambda_dy_c+\lambda_dy_d)}{y_d}.
\end{equation*}
Then calculating $P_{d,y_d}$ is equivalent to solving the following optimization problem:

\begin{eqnarray*}
  && \max_{q}~\mathbb{E}[\exp\{-\frac{s}{q}\}] \\
  &&{\rm s.t.}~~~ \mathbb{E}[q]=1\\
  && ~~~~~~~0\leq q\leq \frac{P_{d,max}^\delta}{y_d}.
\end{eqnarray*}
Set the CDF of $q$ as $F_{q}(x)$, according to Lemma \ref{max}, we have the solution:
\begin{eqnarray*}
F_{q}(x)=
\begin{cases}
\mathbbm{1}_{[1,\infty)}(x) & \text{if} \phantom{aaa} s\leq 1,\\
(1-\frac{1}{s})\mathbbm{1}_{[0,s)}(x)+\mathbbm{1}_{[s,\infty)}(x) & \text{if} \phantom{aaa} 1<s\leq\frac{P_{d,max}^\delta}{y_d},\\
(1-\frac{y_d}{P_{d,max}^\delta})\mathbbm{1}_{[0,\frac{P_{d,max}^\delta}{y_d})}(x)+\mathbbm{1}_{[\frac{P_{d,max}^\delta}{y_d},\infty)}(x) & \text{if} \phantom{aaa} s>\frac{P_{d,max}^\delta}{y_d}.
\end{cases}
\end{eqnarray*}
The $P_{d,y_d}$ is given as following:
\begin{eqnarray}\label{pdyd}
P_{d,y_d}=
\begin{cases}
\exp\{-s\} & \text{if} \phantom{aaa} s\leq 1,\\
\frac{1}{s}\exp\{-1\} & \text{if} \phantom{aaa} 1<s\leq\frac{P_{d,max}^\delta}{y_d},\\
\frac{y_d}{P_{d,max}^\delta}\exp\{-\frac{sy_d}{P_{d,max}^\delta}\}  & \text{if} \phantom{aaa} s>\frac{P_{d,max}^\delta}{y_d}.
\end{cases}
\end{eqnarray}
However, noting that,
\begin{eqnarray*}
\frac{1}{s}\exp\{-1\}<\frac{1}{e}\leq1-\epsilon_d && \text{if} \phantom{aaa} 1<s\leq\frac{P_{d,max}^\delta}{y_d},\\
\frac{y_d}{P_{d,max}^\delta}\exp\{-\frac{sy_d}{P_{d,max}^\delta}\}<\exp\{-\frac{sy_d}{P_{d,max}^\delta}\}<\frac{1}{e}\leq1-\epsilon_d && \text{if} \phantom{aaa} s >\frac{P_{d,max}^\delta}{y_d},\\
s\leq1 && \text{if} \phantom{aaa} \exp\{-s_d\}>1-\epsilon_d.
\end{eqnarray*}
That is to say, under $\epsilon_d\leq1-\frac{1}{e}$, $P_{d,y_d}$ is achieved when $P_d$ is a constant.
Then, we have
\begin{eqnarray*}
\mathcal{F}&=&\bigcup_{y_d}\big\{(\lambda_c,\lambda_d)\succeq0:\exp\{-s\}\geq1-\epsilon_d,\lambda_cy_c+\lambda_dy_d \leq \frac{Q_c}{\phi_c}\big\},\label{CPC}\\
& {=}&\bigcup_{y_d}\big\{(\lambda_c,\lambda_d)\succeq0:\lambda_cy_c+\lambda_dy_d \leq y_d\frac{1}{\phi_d}\ln\frac{1}{1-\epsilon_d},\lambda_cy_c+\lambda_dy_d \leq \frac{Q_c}{\phi_c}\big\},\\
&=&\bigcup_{y_d\in[\frac{Q_c\phi_d}{\phi_c\ln\frac{1}{1-\epsilon_d}},\infty)}\big\{(\lambda_c,\lambda_d)\succeq0:\lambda_cy_c+\lambda_dy_d \leq \frac{Q_c}{\phi_c}\big\},\\
&&\bigcup_{y_d\in[0,\frac{Q_c\phi_d}{\phi_c\ln\frac{1}{1-\epsilon_d}})}\big\{(\lambda_c,\lambda_d)\succeq0:\lambda_cy_c+\lambda_dy_d \leq y_d\frac{1}{\phi_d}\ln\frac{1}{1-\epsilon_d}\big\},\\
& \overset{\text{(a)}}{=} &
\begin{cases}
\big\{(\lambda_c,\lambda_d)\succeq0:-\frac{\phi_d\lambda_d}{\ln(1-\epsilon_d)}+\frac{y_c\phi_c\lambda_c}{Q_c} \leq 1\big\} & \mbox{if} \phantom{aaa} P_{d,max}>\bigg(\frac{Q_c\phi_d}{\phi_c\ln\frac{1}{1-\epsilon_d}}\bigg)^{\frac{1}{\delta}},\\
\big\{(\lambda_c,\lambda_d)\succeq0:\lambda_d+\lambda_c\frac{y_c}{P_{d,max}^\delta} \leq \frac{1}{\phi_d}\ln\frac{1}{1-\epsilon_d}\big\} & \mbox{if}\phantom{aaa} P_{d,max}\leq\bigg(\frac{Q_c\phi_d}{\phi_c\ln\frac{1}{1-\epsilon_d}}\bigg)^{\frac{1}{\delta}},
\end{cases}
\end{eqnarray*}
where (a) from the fact indicated by \eqref{CPC} that the region can be reached when $P_d$ is constant.
\end{IEEEproof}

%\section{Proof of Theorem \ref{FR2}}\label{FR2proof}
%
%%\emph{Remark:}\eqref{largerdelta} implies that unlike the independent case, the delta in the dependent case may exceed 1.
%
%Then we can prove the theorem.
%\begin{IEEEproof}
%Set $P_{i,y_c,y_d}=\min_{\mathbb{E}[P_c^\delta]=y_c,\mathbb{E}[P_d^\delta]=y_d}\underline{Pr_i},i\in\{c,d\}$, similar to the proof of Theorem \ref{FR1}, according to \eqref{feasibilityregion} and \eqref{approxconstrain} we have:
%\begin{eqnarray*}
%\mathcal{F}&=&\bigcup_{(y_c,y_d)}\big\{(\lambda_c,\lambda_d)\succeq0:P_{c,y_c,y_d}<\epsilon_c,P_{d,y_c,y_d}<\epsilon_d\big\},
%\end{eqnarray*}
%
%Therefore,
%\begin{eqnarray*}
%\mathcal{F}&=&\bigcup_{(y_c,y_d)}\big\{(\lambda_c,\lambda_d)\succeq0:
%\frac{\lambda_cy_c+\lambda_dy_d}{y_i}<\frac{1}{\psi_i\Gamma(1-\frac{\delta}{2})^2}\ln\frac{1}{1-\epsilon_i},i\in\{c,d\}\big\},\\
%&=&\big\{(\lambda_c,\lambda_d)\succeq0:
%-\frac{\psi_c\Gamma(1-\frac{\delta}{2})^2\lambda_c}{\ln(1-\epsilon_c)}-\frac{\psi_d\Gamma(1-\frac{\delta}{2})^2\lambda_d}{\ln(1-\epsilon_d)}<1\big\}.
%\end{eqnarray*}
%\end{IEEEproof}
\section{Proof of Theorem \ref{lowestpower1}}\label{lowestpower1proof}
\begin{lemma}\label{eq}
For the two problems \eqref{max3}-\eqref{constraine2} and \eqref{max4}-\eqref{constraine3}, they have the same solution.\\
\begin{eqnarray}
&&\min_{y_d\leq R_c}\min_{P_d}\mathbb{E}[P_d]\label{max4}\\
&&{\rm s.t.}~~~ \mathbb{E}\big[\exp\{-\frac{\phi_d(\lambda_cy_c+\lambda_dy_d)}{P_d^\delta}\}\big]=1-\epsilon_d\label{constrainc3}\\
&& ~~~~~~~\mathbb{E}[P_d^\delta]=y_d\label{constraind3}\\
&& ~~~~~~~0\leq P_d\leq P_{d,max}\label{constraine3}
\end{eqnarray}
\end{lemma}
\begin{IEEEproof}
It suffices to prove that when \eqref{max3} is minimized, the equation \eqref{constraind2} or \eqref{constrainc3} holds. Let $P^\star$ minimizes \eqref{max3}. We assume $\Pr(SIR_d\leq\theta_d)  < \epsilon_d$ or $\mathbb{E}[\exp\{-\frac{\phi_d(\lambda_cy_c+\lambda_dy^\star)}{P_d^\delta}\}]>1-\epsilon_d$. Recall Lemma \ref{continuity} and notice that
\begin{eqnarray*}
\lim_{t\to 0^+}\mathbb{E}[\exp\{-\frac{\phi_d(\lambda_cy_c+\lambda_dt^\delta y^\star)}{t^\delta P_d^\delta}\}]=\lim_{t\to 0^+}\mathbb{E}[\exp\{-\frac{\phi_d(\frac{\lambda_cy_c}{t^\delta}+\lambda_dy^\star)}{P_d^\delta}\}]=0,
\end{eqnarray*}
where $y^\star=\mathbb{E}[P^{{\star}^{\delta}}]$.
There exists a $0<t_0<1$ such that
\begin{eqnarray*}
\mathbb{E}[\exp\{-\frac{\phi_d(\lambda_cy_c+\lambda_dt_0^\delta y^\star)}{t_0^\delta P_d^\delta}\}]=\mathbb{E}[\exp\{-\frac{\phi_d(\frac{\lambda_cy_c}{t_0^\delta}+\lambda_dy^\star)}{P_d^\delta}\}]=1-\epsilon_d.
\end{eqnarray*}
Set $P^{\star\star}=t_0P^\star$, then $P^{\star\star}$ satisfies the constrains (\eqref{constrainc2}, \eqref{constraind2}, \eqref{constraine2}) and $\mathbb{E}[P^{\star\star}]<\mathbb{E}[P^\star]$. This contradicts with the assumption that $P^\star$ is the solution. Thus we complete the proof.
\end{IEEEproof}
\begin{lemma}
Set $y_d=y_0\triangleq\frac{\lambda_cy_c}{\frac{1}{\phi_d}\ln\frac{1}{1-\epsilon_d}-\lambda_d}$, then one solution to the optimization problem:
\begin{eqnarray}
&&\min_{P_d}\mathbb{E}[P_d],\nonumber\\
&&{\rm s.t.}~~~\mathbb{E}[\exp\{-\frac{\phi_d(\lambda_cy_c+\lambda_dy_d)}{P_d^\delta}\}]=1-\epsilon_d,\label{constrainf}\\
&& ~~~~~~~\mathbb{E}[P_d^\delta]=y_0,\nonumber\\
&& ~~~~~~~0\leq P_d\leq P_{d,max},\nonumber
\end{eqnarray}
is that $P_{d0}=y_0^\frac{1}{\delta},a.s.$
\end{lemma}
\begin{IEEEproof}
By simple calculation, it is verified that $P_d=y_0^\frac{1}{\delta},a.s.$ satisfies \eqref{constrainf}.
Then according to H\"{o}lder's inequality\cite{probability} and $\delta<1$, we have $\mathbb{E}[P_{d0}^\delta]=\mathbb{E}[P_{d0}^\delta\cdot 1]\leq(\mathbb{E}[P_{d0}^\delta]^{\frac{1}{\delta}})^\delta(\mathbb{E}[1^{\frac{1}{1-\delta}}])^{1-\delta}=(\mathbb{E}[P_{d0}])^\delta.$ When $P_{d0}=y_0^\frac{1}{\delta},a.s.$, the equation holds and $\mathbb{E}[P_{d0}]$ is minimized.
\end{IEEEproof}
%\begin{lemma}\label{eq2}
%One solution to problem
%\begin{eqnarray*}
%&&\min_{y_0\leq y_d<R_c}\min_{P_d}\mathbb{E}P_d,\\
%\mbox{subject to:}&&\nonumber\\
%&&\mathbb{E}\exp\{-\frac{\phi_d(\lambda_cy_c+\lambda_dy_d)}{P_d^\delta}\}=1-\epsilon_d,\\
%&&\mathbb{E}P_d^\delta=y_d,\\
%&&0\leq P_d\leq P_{d,max},
%\end{eqnarray*}
%is that $P_{d0}=y_0^\frac{1}{\delta},a.s.$.
%\end{lemma}
%\begin{IEEEproof}
%It suffices to prove that the solution to the problem:
%\begin{eqnarray}
%&&\min_{P_d}\mathbb{E}P_d,\\
%\mbox{subject to:}&&\nonumber\\
%&&\mathbb{E}\exp\{-\frac{\phi_d(\lambda_cy_c+\lambda_dy_d)}{P_d^\delta}\}=1-\epsilon_d,\label{constrainc4}\\
%&&\mathbb{E}P_d^\delta=y_d>y_0,\label{constraind4}\\
%&&0\leq P_d\leq P_{d,max},\label{constraine4}
%\end{eqnarray}
%is greater than $y_0^\frac{1}{\delta}$.
%In fact, if $P_d$ satisfies the constrains \eqref{constrainc4}-\eqref{constraine4}, then
%\begin{equation*}
%\mathbb{E}P_d\geq(\mathbb{E}P_d^\delta)^\frac{1}{\delta}>y_0^\frac{1}{\delta}=\mathbb{E}P_{d0}.
%\end{equation*}
%\end{IEEEproof}
then we can prove the theorem.
\begin{IEEEproof}
Note that if $(\mathbb{E}[P_d^\delta])^\frac{1}{\delta}>y_0^\frac{1}{\delta}$, then, $\mathbb{E}[P_d]\geq(\mathbb{E}[P_d^\delta])^\frac{1}{\delta}>y_0^\frac{1}{\delta}=\mathbb{E}[P_{d0}].$ Thus,
The optimum to problem
\begin{eqnarray*}
&&\min_{y_0\leq y_d<R_c}\min_{P_d}\mathbb{E}[P_d],\\
&&{\rm s.t.}~~~\mathbb{E}[\exp\{-\frac{\phi_d(\lambda_cy_c+\lambda_dy_d)}{P_d^\delta}\}]=1-\epsilon_d,\\
&& ~~~~~~~\mathbb{E}[P_d^\delta]=y_d,\\
&& ~~~~~~~0\leq P_d\leq P_{d,max},
\end{eqnarray*}
is no less than $y_0^{\frac{1}{\delta}}$. So, it suffices to prove that when $y_d<y_0$, there is no $P_d$ that satisfies \eqref{constrainc3}, \eqref{constraind3} and \eqref{constraine3}.\\
We consider the problem:
\begin{eqnarray*}
&&\min_{P_d}\mathbb{E}[P_d^\delta],\\
&&{\rm s.t.}~~~\mathbb{E}[\exp\{-\frac{\phi_d(\lambda_cy_c+\lambda_dy_d)}{P_d^\delta}\}]=1-\epsilon_d,\\
&& ~~~~~~~\mathbb{E}[P_d^\delta]=y_d,\\
&& ~~~~~~~0\leq P_d\leq P_{d,max}.
\end{eqnarray*}
It is equivalent to find the $y_d$:
\begin{eqnarray*}
&&\min_{P_d} y_d,\\
&&{\rm s.t.}~~~\mathbb{E}[\exp\{-\frac{\phi_d(\lambda_cy_c+\lambda_dy_d)}{P_d^\delta}\}]\geq1-\epsilon_d,\\
&& ~~~~~~~\mathbb{E}[P_d^\delta]=y_d,\\
&& ~~~~~~~0\leq P_d\leq P_{d,max}.
\end{eqnarray*}
Then,
\begin{eqnarray*}
\min y_d&=&\min\big\{y_d:\max\mathbb{E}[\exp\{-\frac{\phi_d(\lambda_cy_c+\lambda_dy_d)}{P_d^\delta}\}]\geq1-\epsilon_d,\mbox{s.t.,}\mathbb{E}[P_d^\delta]=y_d \\&&\mbox{ and } 0\leq P_d\leq P_{d,max} \big\},\\
& \overset{\text{(a)}}{=} & \min\big\{y_d:\exp\{-\frac{\phi_d(\lambda_cy_c+\lambda_dy_d)}{y_d}\}\geq1-\epsilon_d\},\\
&=&\min\big\{y_d:\exp\{-\phi_d(\frac{\lambda_cy_c}{y_d}+\lambda_d)\}\geq1-\epsilon_d\},\\
&=& y_0,
\end{eqnarray*}
where (a) is from Lemma \ref{max} and $\epsilon_d\leq1-\frac{1}{e}$.
Thus, when $y_d<y_0$, there is no $P_d$ that satisfies \eqref{constrainc3}-\eqref{constraine3}.
\end{IEEEproof}

\section{Proof of Theorem \ref{PFR2t}}\label{PFR2tproof}
Before proof the theorem, we need a lemma.
\begin{lemma}\label{min}
If $\xi\geq0$ is a random variable, $y_c>0$ is a constant and $h$ is an exponentially distributed random variable with PDF $e^{-h}$ , consider this optimization problem:
\begin{eqnarray*}
&&\min_{\xi}\mathbb{E}[\xi^{-1}h^{-\delta}],\\
&&{\rm s.t.}~~~\mathbb{E}[\xi]=y.
\end{eqnarray*}
The solution is given by:
\begin{equation*}
\min_\xi\mathbb{E}[\xi^{-1}h^{-\delta}]=\frac{1}{y}\Gamma(1-\frac{\delta}{2})^2 \mbox{, when,}\phantom{a} \xi=\frac{y}{\Gamma{(1-\frac{\delta}{2}})}h^{-\frac{\delta}{2}}.
\end{equation*}
\end{lemma}
\begin{IEEEproof}
According to Cauchy-Schwartz inequality \cite{probability}, we have:
\begin{equation*}
y\mathbb{E}[\xi^{-1}h^{-\delta}]=\mathbb{E}[\xi]\mathbb{E}[\xi^{-1}h^{-\delta}]\geq(\mathbb{E}[h^{-\frac{\delta}{2}}])^2=\Gamma(1-\frac{\delta}{2})^2.
\end{equation*}
Then,
\begin{equation}\label{largerdelta}
\mathbb{E}[\xi^{-1}h^{-\delta}]\geq\frac{1}{y}\Gamma(1-\frac{\delta}{2})^2.
\end{equation}
with equation holding if and only if $\xi=\frac{y}{\Gamma{(1-\frac{\delta}{2}})}h^{-\frac{\delta}{2}}, a.s.$.
\end{IEEEproof}

Then we can prove the theorem.
\begin{IEEEproof}
Set $P_{d,y_c,y_d}=\min_{\mathbb{E}[P_c^\delta]=y_c,\mathbb{E}[P_d^\delta]=y_d}\underline{Pr_d},y_c=\mathbb{E}[P_c^\delta]$.
We first calculate $P_{d,y_c,y_d}$. According to Lemma \ref{min}, we have
\begin{eqnarray*}
P_{d,y_c,y_d}&=& \min_{\mathbb{E}[P_c^\delta]=y_c,\mathbb{E}[P_d^\delta]=y_d}\bigg(
1-\exp\big\{-\psi_d(\lambda_c\mathbb{E}[P_c^\delta]+\lambda_d\mathbb{E}[P_d^\delta])\mathbb{E}[(h_dP_d)^{-\delta}]\big\}\bigg),\\
&=& 1-\exp\big\{-\psi_d(\lambda_cy_c+\lambda_dy_d)\min_{\mathbb{E}[P_c^\delta]=y_c,\mathbb{E}[P_d^\delta]=y_d}(\mathbb{E}[(h_dP_d)^{-\delta}])\big\},\\
&=&
1-\exp\big\{-\frac{\psi_d(\lambda_cy_c+\lambda_dy_d)}{y_d}\Gamma(1-\frac{\delta}{2})^2\big\}.
\end{eqnarray*}
\begin{eqnarray*}
\mathcal{F}&=&\bigcup_{y_d}\big\{(\lambda_c,\lambda_d)\succeq0:\underline{Pr}_c\leq\epsilon_c,P_{d,y_c,y_d}\leq\epsilon_d\big\},\\
&=&\bigcup_{y_d}\big\{(\lambda_c,\lambda_d)\succeq0:\lambda_cy_c+\lambda_dy_d\leq\frac{y_d}{\psi_d\Gamma(1-\frac{\delta}{2})^2}\ln\frac{1}{1-\epsilon_d},\\
&&\phantom{\bigcup_{y_d}\big\{(\lambda_c,\lambda_d)\succeq0:a} \lambda_cy_c+\lambda_dy_d\leq\psi_c^{-1}\mathbb{E}[P_c^{-\delta}h_c^{-\delta}]^{-1}\ln\frac{1}{1-\epsilon_c}\big\},\\
&=&\big\{(\lambda_c,\lambda_d)\succeq0:\lambda_cy_c+\lambda_d\frac{\psi_d\Gamma(1-\frac{\delta}{2})^2\ln(1-\epsilon_c)}{\psi_c\mathbb{E}[P_c^{-\delta}h_c^{-\delta}]\ln(1-\epsilon_d)}\leq\psi_c^{-1}\mathbb{E}[P_c^{-\delta}h_c^{-\delta}]^{-1}\ln\frac{1}{1-\epsilon_c}\big\}.
\end{eqnarray*}
\end{IEEEproof}

\section{Proof of Lemma \ref{function}}\label{functionproof}

\begin{IEEEproof}
If $\delta\leq1$, set $P^{\star\star}(h)=(\mathbb{E}[(P^{\star})^{\delta}|h])^{\frac{1}{\delta}}$  \footnote{Strictly, it should be written as $P^{\star\star}(h_0)=\mathbb{E}[(P^{\star})^{\delta}|h=h_0]^{\frac{1}{\delta}}$, to simplify the notation we use $P^{\star\star}(h)=\mathbb{E}[(P^{\star})^{\delta}|h]^{\frac{1}{\delta}}$}, so given $h,P^{\star\star}$ is a deterministic function \cite{probabilityross}. According to Jensen's inequality for conditional expectation \cite{probability},
\begin{eqnarray*}
P^{\star\star}(h)&=&(\mathbb{E}[{P^{\star}}^{\delta}|h])^{\frac{1}{\delta}}\geq0\\
\mathbb{E}[(P^{\star\star})^\delta]=\mathbb{E}[\mathbb{E}[(P^{\star})^{\delta}|h]]=\mathbb{E}[(P^{\star})^\delta]&\leq& \bigg(-\frac{\ln(1-\epsilon_c)}{\mathbb{E}[P_c^{-\delta}h^{-\delta}]\psi_c}-\lambda_cy_c\bigg)\frac{1}{\lambda_d}\\
(\lambda_cy_c+\lambda_d\mathbb{E}[{({P^{\star\star}})^\delta}]\mathbb{E}[({P^{\star\star}})^{-\delta}h^{-\delta}]
&=&(\lambda_cy_c+\lambda_d\mathbb{E}[{{P^{\star}}^\delta}])\mathbb{E}\big[(\mathbb{E}[{P^\star}^\delta|h])^{-1}h^{-\delta}\big]\\
&=&(\lambda_cy_c+\lambda_d\mathbb{E}[{{P^{\star}}^\delta}])\mathbb{E}\big[(\mathbb{E}[{P^\star}^\delta h^{\delta}|h])^{-1}\big]\\
&\leq&(\lambda_cy_c+\lambda_d\mathbb{E}[{{P^{\star}}^\delta}])\mathbb{E}\big[\mathbb{E}[{P^\star}^{-\delta}h^{-\delta}|h]\big]\\
&=&(\lambda_cy_c+\lambda_d\mathbb{E}[{{P^{\star}}^\delta}])\mathbb{E}\big[{P^\star}^{-\delta}h^{-\delta}\big]\\
&\leq&\frac{1}{\psi_d}\ln\frac{1}{1-\epsilon_d}
\end{eqnarray*}
Then \eqref{constrainc5}, \eqref{constraind5} and \eqref{constraine5} are satisfied.
Meanwhile,
\begin{equation*}
\mathbb{E}[P^{\star\star}]=\mathbb{E}[(\mathbb{E}[(P^{\star})^{\delta}|h])^{\frac{1}{\delta}}]\leq\mathbb{E}[\mathbb{E}[(P^{\star})|h]]=\mathbb{E}[P^{\star}]
\end{equation*}
%Then, $P^{\star\star}$ satisfy \eqref{constrainc5}, \eqref{constraind5} and \eqref{constraine5}.
If $\delta>1$, set $P^{\star\star}(h)=(\mathbb{E}[(P^{\star})^{\frac{1}{\delta}}|h])^{\delta}$, so given $h,P^{\star\star}$ is a deterministic function \cite{probabilityross}. According to Jensen's inequality for conditional expectation \cite{probability},
\begin{eqnarray*}
P^{\star\star}&=&(\mathbb{E}[(P^{\star})^{\frac{1}{\delta}}|h])^{\delta}\geq0\\
\mathbb{E}[(P^{\star\star})^\delta]=\mathbb{E}[(\mathbb{E}[(P^{\star})^{\frac{1}{\delta}}|h])^{\delta^2}]\leq\mathbb{E}[\mathbb{E}[(P^{\star})^{\delta}|h]]&\leq& \bigg(-\frac{\ln(1-\epsilon_c)}{\mathbb{E}[P_c^{-\delta}h^{-\delta}]\psi_c}-\lambda_cy_c\bigg)\frac{1}{\lambda_d}\\
(\lambda_cy_c+\lambda_d\mathbb{E}[({P^{\star\star}})^\delta])\mathbb{E}[({P^{\star\star}})^{-\delta}h^{-\delta}]
&\leq&(\lambda_cy_c+\lambda_d\mathbb{E}[{P^{\star}}^\delta])\mathbb{E}\big[(\mathbb{E}[{P^\star}^{\frac{1}{\delta}}|h])^{-\delta^2}h^{-\delta}\big]\\
&\leq&(\lambda_cy_c+\lambda_d\mathbb{E}[{P^{\star}}^\delta])\mathbb{E}\big[\mathbb{E}[{P^\star}^{-\delta}|h]h^{-\delta}\big]\\
&=&(\lambda_cy_c+\lambda_d\mathbb{E}[{P^{\star}}^\delta])\mathbb{E}\big[{P^\star}^{-\delta}h^{-\delta}\big]\\
&\leq&\frac{1}{\psi_d}\ln\frac{1}{1-\epsilon_d}
\end{eqnarray*}
Then \eqref{constrainc5}, \eqref{constraind5} and \eqref{constraine5} are satisfied.
Meanwhile,
\begin{equation*}
\mathbb{E}[P^{\star\star}]=\mathbb{E}[(\mathbb{E}[(P^{\star})^{\frac{1}{\delta}}|h])^{\delta}]\leq\mathbb{E}[\mathbb{E}[(P^{\star})|h]]=\mathbb{E}[P^{\star}]
\end{equation*}
\end{IEEEproof}




%\bibliographystyle{IEEEtran}

\bibliographystyle{IEEEtran}
\bibliography{IEEEabrv,Reference}
\end{document}